\date{}
\documentclass[aps,pra,showpacs,twocolumn]{revtex4}
\usepackage{graphicx,psfrag,amsmath,amssymb,amsfonts,latexsym,color,dcolumn}

\begin{document}

\title{Geometric phases under the presence of a composite environment}

\author{Paula I. Villar$^{1}$ \footnote{paula@df.uba.ar}}
\author{Fernando C. Lombardo$^1$ \footnote{lombardo@df.uba.ar}}

 \affiliation{$^1$ Departamento de F\'\i sica {\it Juan Jos\'e
 Giambiagi}, FCEyN UBA, Facultad de Ciencias Exactas y Naturales,
 Ciudad Universitaria, Pabell\' on I, 1428 Buenos Aires, Argentina }

\date{today}

\begin{abstract}
We compute the geometric phase for a spin-1/2 particle under the
presence of a composite environment, composed of an external bath
(modeled by an infinite set of harmonic oscillators) and another
spin-1/2 particle. We consider both cases: an initial entanglement
between the spin-1/2 particles and an initial product state in order to see if the
initial entanglement has an enhancement effect on the geometric
phase of one of the spins. We follow the nonunitary evolution of the 
reduced density matrix and evaluate the geometric phase for a single 
two-level system. We also show that the initial entanglement enhances the 
sturdiness of the geometric phase under the presence of an external composite 
environment.
\end{abstract}


\maketitle

\section{Introduction}

It is widely known that a system can retain the information of its motion when
it undergoes a cyclic evolution, in the form of a geometric
phase (GP). This  was first put forward by Pancharatnam in optics
\cite{Pancharatman} and later studied explicitly by Berry
in a general quantal system \cite{Berry}. Since then, great
progress has been achieved in this field as it became clear that geometric
phases had important consequences for quantum systems.  The original notion of
Berry phase has been extended to the case of non adiabatic evolutions \cite{Anandan}.
As an important evolvement, the application of the geometric
phase has been proposed in many fields, such as
the geometric quantum computation. Due to its global properties,
the GP is propitious
to  construct  fault tolerant  quantum  gates.
The study of the GP was soon extended to
open quantum systems. In this context, many authors have
analyzed the correction to the GP under the
influence of an external environment using different approaches
\cite{Whitney, Carollo, dechiara, tong, PRA, nos, pau, Sanders}.
The interest on geometric phases has lately reached composite systems
such as bipartite systems. In a previous article \cite{PRA81}, we have computed the
GP for the bipartite system under the influence of a nonunitary
evolution, induced by a bosonic or fermionic environment. There, we have 
shown that entanglement plays an important role in the robustness of the 
GP for open systems. In \cite{Tong81},
it was shown a particular situation in which an initial separable state remains
separable so that the GP of the system is always
equal to the sum of the geometric phases of its subsystems under an unitary evolution.

The presence of an environment can destroy all the traces of the
quantumness of a system. All real world quantum systems interact
with their surrounding environment to a greater or lesser extent. As
the quantum system is in interaction with an environment defined as
any degrees of freedom coupled to the system which can entangle its
states, a degradation of pure states into mixtures takes place. No
matter how weak the coupling that prevents the system from being
isolated, the evolution of an open quantum system is eventually
plagued by nonunitary features like decoherence and dissipation.
Decoherence, in particular, is a quantum effect whereby the system
loses its ability to exhibit coherent behavior. Nowadays,
decoherence stands as a serious obstacle in quantum information
processing.

In this context, there are many efforts to find a GP 
that ``survives'' the presence of the environment. What's more, the 
GP for a state under nonunitary evolution has not yet
been directly measured. Recently, using a tomographic approach, the
GP was measured for a  qubit coupled to a critical environment using
a nuclear magnetic resonance (NMR) quantum simulator
\cite{cucchiettiprl}.

In this paper,  we  study the  geometric phase 
(GP) of a qubit (spin 1) in a noisy  `composite" environment. That means, we 
start with a two-qubit-state (spin 1 and spin 2) 
coupled to an external reservoir and then, we  focus on 
only one qubit (by tracing out the other spin 2) in order to study the geometric phase 
(GP) of the resulting spin 1 (our subsystem). In the end, the model is different to other
preexisting models, such as for example, 
two-qubits in a noisy environment or/and a solely spin in a noisy
environment.  The reason for 
choosing this model is twofold. On one side, it is the natural sequel to other previous studies
of the GP in open quantum systems and  has interesting features that 
differentiate this model from the others studied previously. On the other side, 
by taking into account new results about the existence 
of an environmentally-induced GP for quantum states \cite{sqjovist10}, 
we present a new insight on the effect that the degree
of entanglement has on the geometric phase of a particle of
spin-1/2 and, a possible way to measure it using NMR quantum simulators. 
Following this line, we shall start by describing the
model and computing the geometric phase for a general environment in
Section \ref{II}. In Section \ref{III}, we shall focus on a bosonic
environment writing explicitly the decoherence factors involved.
We shall consider different situations for the initial
state of the bipartite: either entangled (Sec. \ref{entangled}) or a
product state (Sec. \ref{product}). Further, in Section \ref{nodisipation}
we shall study the differences that arise in the former computations
if one of the spin-1/2 particles is not coupled to the external
environment. Finally, we conclude our results in Section \ref{IV}.

\section{Decoherence induced on one two-level
system due to a composite environment}
\label{II}

We shall consider a bipartite system (as in \cite{PRA81}), that is
to say, two interacting two-level systems, both coupled to an
external reservoir. We shall study a model
where the Hamiltonians that describe the complete evolution are
given by $H = H_S + H_I + H_B$ where the system's and interaction
Hamiltonians are respectively defined as
\begin{eqnarray}
H_S &=& \frac{\hbar \Omega_1}{2} \sigma_z^1 + \frac{\hbar
\Omega_2}{2} \sigma_z^2 + \chi~ \sigma_z^1 \otimes \sigma_z^2\\
H_I &=& \sigma_z^1 \otimes \sum_{n=1}^N \lambda_{1n} q_n +
\sigma_z^2 \otimes \sum_{n=1}^N \lambda_{2n} q_n.
\end{eqnarray}
The environment's Hamiltonian $H_B$ will be left without definition
until we are faced with a particular example.

One approach to know the nonunitary features induced by the
environment in the bipartite subsystem is by tracing out the degrees
of freedom of the environment as suggested in Ref.
\cite{leshouches}. For a general arbitrary two-qubit initial state
\begin{equation}
| \Psi(0) \rangle = \alpha |0 0 \rangle + \beta | 0 1 \rangle  + \zeta
|1 0 \rangle + \delta | 1 1 \rangle,
\label{estadogen}
\end{equation}
 the reduced
density matrix takes the form:

\begin{widetext}
\begin{equation}
\rho_{\rm r}(t)=\left(\begin{array}{cccc} |\alpha|^2 & \alpha
\beta^* e^{-i(2 \chi +\Omega_2)t} F_{12} &
\alpha \zeta^*  e^{-i(2 \chi +\Omega_1)t} F_{13} & \alpha \delta^*
e^{-i(\Omega_1+\Omega_2)t} F_{14} \\
\beta \alpha^*  e^{i(2 \chi +\Omega_2)t} F_{12}^*  &
 |\beta|^2  & \beta \zeta^*
e^{-i(\Omega_1-\Omega_2)t} F_{23} &
\beta \delta^* e^{-i(\Omega_1 -2 \chi)t} F_{24}\\
\zeta \alpha^* e^{i(2 \chi +\Omega_1)t} F_{13}^* &
\zeta \beta^* e^{i(\Omega_1-\Omega_2)t} F_{23}^* & |\zeta|^2  &
 \zeta \delta^* e^{-i(\Omega_2-2 \chi)t} F_{34} \\
\delta \alpha^* e^{i(\Omega_1+\Omega_2)t} F_{14}^* &
\delta \beta^* e^{i(\Omega_1 -2 \chi)t} F_{24}^*
& \delta \zeta^* e^{i(\Omega_2-2 \chi)t} F_{34}^*  & |\delta|^2
\end{array}\right).
\label{rhogen}
\end{equation}
\end{widetext}

By $F_{i}$, with $i=12,13,14,23,24,34$, we are implying the
decoherence and dissipation factors that appear due to the presence
of the environment in the reduced density
matrix. These factors are obtained from a noise and
dissipation kernels similarly as in the familiar spin-boson treatment. 
The master equation for the reduced density matrix of a bipartite system coupled to
a bosonic environment (Eq.\ref{rhogen}) and the calculation of these terms can be read
in the Appendix A of \cite{PRA81}.
However, we are not interested in the explicit form of these 
coefficients right now, since we
want to focus on  the geometric phase acquired by only one spin of the
bipartite subsystem. For this, we need to know the dynamics of this
only spin ruled by its own Hamiltonian and the presence of the
composite environment: the reservoir plus the other spin-1/2 particle of
the subsystem. It is important to note that both spin-1/2 particles
interact by the coupling constant $\chi$ defined in $H_S$ and each
of them interact as well with the external environment by the
coupling constants $\lambda_{1n}$ and $\lambda_{2n}$, respectively.
We could further define the coefficients of the initial bipartite state
so as to be able to interpret the results in terms of the known Bloch ball
(we have not done so before because of the difficulty of writing Eq.(\ref{rhogen})).
Therefore, we can assume that
\begin{eqnarray}
 \alpha &=& \sqrt{\lambda_0} \cos(\theta_0/2) \nonumber \\
 \beta &=& -\sqrt{\lambda_1} \sin(\theta_0/2) \nonumber \\
 \zeta &=& \sqrt{\lambda_0} \sin(\theta_0/2) \nonumber \\
 \delta &=& \sqrt{\lambda_1} \cos(\theta_0/2).
\label{estadoinicial}
\end{eqnarray}
The normalization of this state implies $\lambda_1 = 1-\lambda_0$.
The concurrence of this state is ${\cal C}= 2 \sqrt{\lambda_0
\lambda_1}$. Hence, we see that when $\lambda_0 = \lambda_1 =1/2$ we
have a maximally entangled state (MES). The angle $\theta_0$ has
here a geometric interpretation: coordinates $X= (1-2 \lambda_0)
\sin(\theta_0)$, $Y=0$ and $Z=(2 \lambda_0 -1) \cos(\theta_0)$ of
the state vector in the Bloch ball are given, using the results
obtained in \cite{Milman}. The radius of the Bloch ball is $R=|2
\lambda_0-1|$ and all states lying on its surface have the same
degree of purity (and come from two qubit states having the same
amount of entanglement). The angle $\theta_0$ is the one that makes
the density matrix vector with the z axis. 

In order to know the dynamics of only one spin we need to trace out 
 the degrees of freedom corresponding to the spin 2. After doing so,
 in this familiar notation, the reduced density matrix $\tilde{\rho}_{\rm r}(t)$ for
the dynamics of spin 1 is
\begin{widetext}
\begin{equation}
\tilde{\rho}_{\rm r}(t)=\left(\begin{array}{cc} (\lambda_0 -1/2)
\cos\theta_0 + 1/2
 & 1/2 \sin\theta_0 e^{-i \Omega_1 t} \Gamma(t)  \\
 1/2 \sin\theta_0 e^{i \Omega_1 t}  \Gamma(t)^*
&  (1/2 -\lambda_0) \cos\theta_0 + 1/2\\
\end{array}\right).
\label{rho1gen}
\end{equation}
\end{widetext}

It is important to note that the $\Gamma(t)$ factor includes a real
decaying decoherence factor as well as an imaginary dissipation
term, both induced by the external parties to our subsystem spin 1.
Therefore, we can write this coefficient as $\Gamma(t) = r(t) \exp(i
\vartheta(t))$, where $r(t)$ is the modulus of the complex number and
$\vartheta(t)$ its argument.

The GP for a  state under nonunitary evolution
has been defined in \cite{tong} as
\begin{eqnarray} \Phi_g & = &
{\rm arg}\bigg\{\sum_k \sqrt{ \varepsilon_k (0) \varepsilon_k (\tau)}
\langle\Psi_k(0)|\Psi_k(\tau)\rangle \times \nonumber \\
& & e^{-\int_0^{\tau} dt \langle\Psi_k|
\frac{\partial}{\partial t}| {\Psi_k}\rangle}\bigg\}, \label{fasegeo}
\end{eqnarray}
where $\varepsilon_k(t)$ are the eigenvalues and
 $|\Psi_k\rangle$ the eigenstates of the reduced density matrix
$\tilde{\rho_{\rm r}}$, and $\tau$ is a time ($\tau >0$) at which we 
study our system. Since we are working under the weak coupling limit, it is useful to
consider a quasi cyclic path ${\cal P}:~ t~
\epsilon~ [0,\tau]$, with $\tau= 2 \pi/ \Omega_1$ ($\Omega_1$ is the
system's characteristic frequency).
As we can see from the GP definition, in order to compute the phase
gained by the system, it is crucial to know the system's dynamics at
all times. This is why this is known as the kinematic approach in
contrast to other approaches for GP in open quantum systems
\cite{Whitney, Carollo, dechiara}. The central result of Eq.
(\ref{fasegeo}) is to extract from the global phase gain acquired
during the evolution, by a proper choice of the ``parallel transport
 condition'', the purification
independent part which can be termed a geometric phase because it is 
gauge invariant and reduces
to the known results in the limit of an unitary evolution.

The eigenvectors and eigenvalues of Eq.(\ref{rho1gen}) are easily
computed as
\begin{eqnarray}
 \varepsilon_{\pm}&=& \pm\frac{1}{2} \sqrt{\cos^2\theta_0
+ r^2 \sin^2\theta_0 + 4 \lambda_0 (\lambda_0 -1)\cos^2\theta_0}
\nonumber \\
&+&  \frac{1}{2}\nonumber \\
|\upsilon_{\pm} \rangle &=& \frac{e^{-i \Omega_1 t + i \vartheta(t)} r(t)
\sin\theta_0}{\sqrt{r^2 \sin^2\theta_0 + [(2 \varepsilon_{\pm}-1)
+ (1-2 \lambda_0) \cos\theta_0]^2}} |0 \rangle \nonumber \\
&+& \frac{[(2 \varepsilon_{\pm}-1)
+ (1-2 \lambda_0)]}{\sqrt{r^2 \sin^2\theta_0 + [(2 \varepsilon_{\pm}-1)
+ (1-2 \lambda_0)\cos\theta_0]^2}} |1\rangle \nonumber
\end{eqnarray}

Unfamiliarly to all examples done before, this time, for an arbitrary
initial state, we see that $\varepsilon_+(0)\neq 1$ and $\varepsilon_-(0)\neq 0$.
Then, there is contribution to the GP coming from both eigenvalues.
The only real factors in Eq.(\ref{fasegeo}) are $\alpha=\sqrt{\varepsilon_+(0)
\varepsilon_{+}(\tau)}$ and $\beta=\sqrt{\varepsilon_-(0)
\varepsilon_{-}(\tau)}$. After some algebra, we obtain the geometric
phase gained by spin 1 under this non-unitary evolution in a time $\tau$
\begin{widetext}
\begin{eqnarray}
\Phi_g &=& \mathrm{arg}\bigg( \alpha r_+(\tau) e^{i \varphi_+} e^{i \varphi } +
\beta r_-(\tau) e^{i \varphi_-} e^{i {\tilde{\varphi}}}
\bigg)
\label{calculofasegen}
\end{eqnarray}
\end{widetext}

where we have defined
\begin{eqnarray}
  r_+(\tau) e^{i \varphi_+} &=& \langle \upsilon_+(0)|\upsilon_+(\tau) \rangle, \\
 r_-(\tau) e^{i \varphi_-} &=& \langle \upsilon_-(0)|\upsilon_-(\tau) \rangle,   \\
 \varphi &=& \int_0^{\tau} (\Omega_1 + \frac{d \vartheta}{d t}) \cos^2\theta_+(t) ~dt \label{fi},\\
 \tilde{\varphi} &=& \int_0^{\tau} (\Omega_1 + \frac{d \vartheta}{d t}) \cos^2\theta_-(t) ~dt \label{tildefi}.
\end{eqnarray}

We have also used
\begin{eqnarray}
\cos\theta_{\pm}(t) &=& \frac{r(t) \sin\theta_0}{\sqrt{r^2 \sin^2\theta_0 +
[(2 \varepsilon_{\pm}-1) +(1-2 \lambda_0)\cos\theta_0]^2}} \nonumber
\end{eqnarray}
so as to be able to write the eigenvector in an easier way:
\begin{eqnarray}
 |\upsilon_{\pm} \rangle &=& e^{-i \Omega_1 t + i \vartheta(t)} \cos\theta_{\pm}(t) |0\rangle
+ \sin\theta_{\pm}(t) |1\rangle
\end{eqnarray}

\section{Bosonic environment}
\label{III}

In this Section, we investigate the effect of a bosonic environment
coupled to the composite system of two spin-1/2 particles, whose
Hamiltonian is defined as
\begin{equation}
 H_B =\sum_{n=1}^N \hbar \omega_n a_n^{\dagger} a_n.
\end{equation}
One assumption we shall make
is that the spectral density of the bosonic environment $J(\omega)$
is a reasonably smooth function of $\omega$, and
that is of the form $\omega^n$ up to some cutoff frequency $\Lambda$ that
may be large compared to $\Omega_1$ and $\Omega_2$. The spectral
density function can be  written as $J(\omega)= {\gamma_0}/4
~\omega^n \Lambda^{n-1} e^{-\omega/\Lambda} $, where
 dimensionless $\gamma_{01} \sim \lambda_1^2$ and $\gamma_{02} \sim \lambda_2^2$ \cite{Leggett}.
We can  consider an ohmic spectral density for
such an environment, particularly one that goes as $J(\omega) \sim
\omega$. In that case,
the $\Gamma(t)$ factor that appears due to the
tracing out of the degrees of freedom of the environment and the
degrees of freedom of the spin 2, is
\begin{equation}
 \Gamma(t) =  e^{-2 \gamma_{0} \log(1+\Lambda^2 t^2)}
 \big(  (2 \lambda_0 -1) \cos\Omega_R t
- i  \sin\Omega_R t \big)
\label{decoB1}
\end{equation}
with $\Omega_R = (2 \chi-\gamma_{0} \Lambda)$.
We have so far considered  $\gamma_{01}=\gamma_{02}=\gamma_0$ 
for the sake of simplicity.

\subsection{Numerical results for an initial entangled state}
\label{entangled}

\begin{figure}[h!t]
\begin{center}
\includegraphics[width=9.5cm]{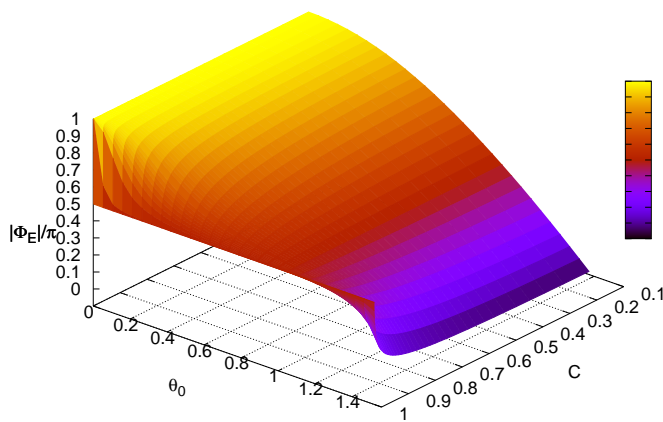}
 \caption{(Color online). Geometric phase $\Phi_E$ as a function of the concurrence 
${\cal C}$ and the angle $\theta_0$ for the uncoupled case $\chi= 0$ 
and $\gamma_0=0$. The phase is measured in units of $\pi$.}
\label{Fig0}
\end{center}
 \end{figure}

We shall start by considering the isolated spin, initially set up in a bipartite
state, but with no external interaction, that is to say
$\gamma_{0}=0$ and $\chi=0$. Hence, in this uncoupled situation of the bipartite
initial state and after tracing out spin number 2, there is a decoherence
factor $\Gamma(t)= (2 \lambda_0-1)$ for spin 1. This means that the initial degree of entanglement
will affect the dynamics through the decoherence factor.  It is also important to note,
that when $\lambda_0=1/2$, we have a maximal entangled state (MES), and
the decoherence factor is $\Gamma=0$, killing all coherences in Eq.(\ref{rho1gen})
as expected.
 In Fig.\ref{Fig0} we can see
the behavior of the geometric phase (that we call $\Phi_E$ in the case of an 
initially entangled state) in the case spin 1 is
completely isolated but initially belonged to an entangled bipartite state.
 It is easy to see that it does not correspond
to the geometric phase of a unitary single spin (i.e.
$\Phi^U=\pi(1-\cos\theta_0)$) due to the initial entanglement. However, it can serve as a
reference for studying either the influence of the environment or
the entanglement of the bipartite state. One important feature to
note is the monotonic 
behavior of the geometric phase in this case.

\begin{figure}[h!t]
\begin{center}
\includegraphics[width=8.cm]{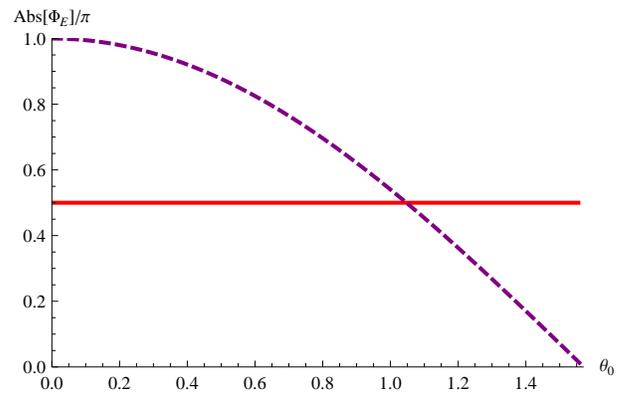}
 \caption{(Color online). Geometric phase of one spin initially
 set up in an entangled bipartite state for the isolated
 case $\gamma_0=0=\chi$ as a function of $\theta_0$. Red curve for ${\cal C}=1$,
 and purple curve for ${\cal C}=0.06$. The phase is measured in units of $\pi$.}
\label{Fig1}
\end{center}
 \end{figure}

If the degree of entanglement is $\lambda_0=1/2$, we have
a MES and the concurrence is ${\cal{C}}=1$, for all values of $\theta_0$.
In Fig.\ref{Fig1} we present  the geometric phase
for two particular initial entangled states of different concurrences for a
better understanding of the behavior. In this figure, we present a 
solid red curve for maximum entanglement and a dashed purple one
for a very low initial entanglement. We can note that for a spin
initially set up in a 
MES the GP is $\pi/2$.
On the other side, we can see that for an ``almost" initial separable state
(${\cal C} \sim 0$),
the value of the geometric phase for only one spin depends strongly
on the value of $\theta_0$ (purple dashed line in Fig.\ref{Fig1}).
 In any case, an initially separable (product) state shall be further discussed in 
 a following Section.

Let's recall that for a bipartite state, we can not longer use
the Bloch sphere to seek a geometric representation of that state.
In \cite{milman} it has been shown that a geometric representation
of a bipartite state can be obtained by using a Bloch ball and a SO3
sphere. Therefore,  the total phase gained by a state is a
combination of not only the dynamical and geometrical phase, but
also the topological phase. Similarly to one qubit states, MES also
gain a total phase of $\pi$ (or n$\pi$) under a cyclic evolution.
However, this phase is of topological origin. It is already known
that for a qubit the total phase gained is $\pi$ (or n$\pi$) and it
is due to a combination of the dynamical phase and the GP.  
Now, we are obtaining like the ``partial" phase of that entangled
state, i.e.  the GP of one spin-1/2 particle after tracing out the ``external''
degree of freedom of the other spin, both initially set up in an
isolated bipartite state. As a MES can not be represented
in the Bloch Sphere, i.e. it is reduced to the central point of it,
the eigenvalues of the reduced density matrix for $\lambda_0=1/2$
are degenerate. By using the definition of the GP for the degenerate case
proposed also in \cite{tong}, which in this case reduces mainly to the same 
formulation multiplied by the degeneracy of the eigenvalues, it is easy to obtain the value of the GP
for this case, namely $\pi/2$. Another case that can be easily analyzed is
when $\theta_0=0$. Hence, the initial state reduces to
$|\Psi(0)\rangle = \sqrt{\lambda_0} |0 0 \rangle +
\sqrt{1-\lambda_0} |1 1 \rangle$, and it is easy to see that the
$\cos\theta_{\pm}(t)=\pi$. Then, doing some algebra, we obtain that
the $\Phi_E = \pi$ for all $\lambda_0 \neq 1/2$.
On the other hand, if we have $\theta_0=\pi$, which is equivalent to having the initial
state $|\Psi(0)\rangle = -\sqrt{1-\lambda_0} |0 1 \rangle +
\sqrt{\lambda_0} |1 0 \rangle$, the GP for a spin is zero for all $\lambda_0 \neq 1/2$.
These two initial states mentioned are the Werner states studied in \cite{PRA81}.
 Thus, we see that the GP of a spin initially entangled in a bipartite state, 
 yields a similar geometric phase than the one obtained for the whole
bipartite isolated system.\\

\begin{figure}[h!t]
\begin{center}
\includegraphics[width=6.cm]{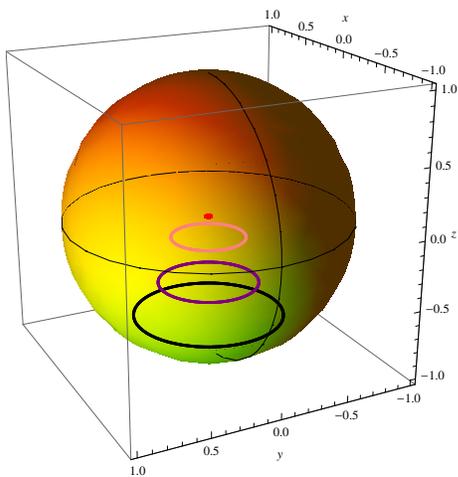}
 \caption{(Color online). The evolution of the system can be illustrated by the path traversed in the
 Bloch sphere in the case that $\gamma_0=0$ and $\chi=0$. Red curve for ${\cal {C}}=1$, pink curve for ${\cal {C}}=0.95$, 
purple curve for ${\cal {C}}=0.8$, and black curve ${\cal {C}}=0.43$. Time is measured in
units of $\Omega_1$.}
 \label{Fig2}
\end{center}
 \end{figure}
As we are focusing on the geometric phase of one spin, we can find the
path traversed in the Bloch sphere by studying the several components
of the reduced density matrix $\tilde{\rho}_r$. The three-dimensional coordinates in the
 Bloch sphere are $x={\tilde {\rho}}_{r_{12}}+{\tilde{\rho}}_{r_{21}}$, $y= i
 ({\tilde{\rho}}_{r_{12}}-{\tilde{\rho}}_{r_{21}})$ and $z={\tilde{\rho}}_{r_{11}}-
{\tilde{\rho}}_{r_{22}}$. By the use of the Eq.(\ref{rho1gen}), it is easy to generate the trajectories of Fig.\ref{Fig2}.
Thus, we can see a point for a MES, and bigger circles corresponding to the trajectories
of the states with smaller degrees of initial entanglement, i.e smaller values of ${\cal C}$,
when $\chi=0$ and $\gamma_0=0$.\\

\begin{figure}[h!t]
\begin{center}
\includegraphics[width=9.5cm]{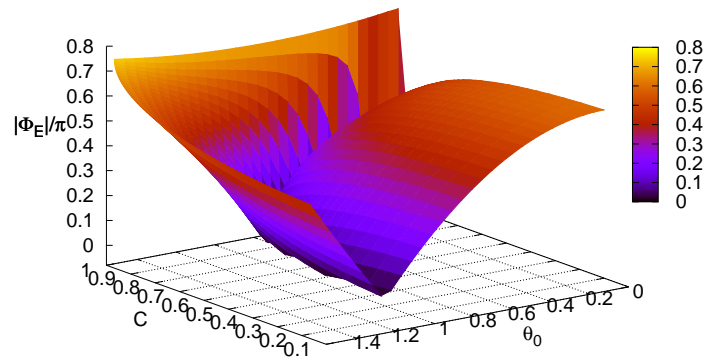}
 \caption{(Color online). Geometric phase $\Phi_E$ as a function of the initial
 degree the entanglement through ${\cal C}$ and the angle $\theta_0$ for $\gamma_0=0$ and $\chi=0.1\Omega_1$.
The phase is normalized by $\pi$.}
\label{Fig3}
\end{center}
 \end{figure}

If we consider that there is no bosonic environment ($\gamma_0=0$) 
but the two spin-1/2
particles are coupled through the $\chi$ constant, then the decoherence
factor is:
\begin{eqnarray}
\Gamma_{\chi}(t) &=&  (2 \lambda_0 -1) \cos(2 \chi t)- i\sin(2 \chi t), 
\nonumber
\end{eqnarray}
which explicitly shows that the coupling between both particles
not only affects the dynamics through the time-dependent decoherence complex
factor but the geometric phase as well. 
In all cases, except for $\lambda_0 = 1/2$, the degree of
entanglement and coupling constant $\chi$ act as a source of noise and dissipation (through its
real and imaginary parts) for the system particle. In
Fig.\ref{Fig3} we show the behavior of the geometric phase of the
spin coupled to the other spin.
 One important feature that can be noted in Fig.\ref{Fig3} is that
the monotonic behavior is broken. The
main reason for this, is the contribution of both eigenenergies
$\varphi$ and $\tilde \varphi$ in Eqs. (\ref{fi}) and
(\ref{tildefi}) and the presence of a dissipation factor. Another interesting feature is that for a
maximally entangled state (i.e. $\lambda_0=1/2$, ${\cal C}=1$)
the geometric phase is similar to that of the isolated case ($40\%$ bigger).
The presence of the valley is due to the intrinsic dynamics between
the spin introduced by the coupling constant $\chi$.\\

\begin{figure}[h!t]
\begin{center}
\includegraphics[width=9.cm]{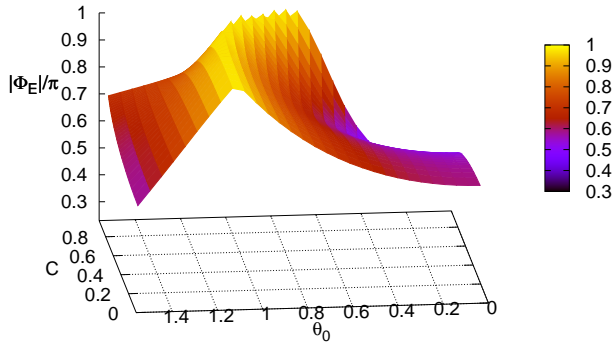}
 \caption{(Color online). Geometric phase $\Phi_E$ as a function of $\cal C$
and $\theta_0$. The value of the geometric
phase is measured in units of $\pi$.  Parameters used: $\chi=0.1\Omega_1$, $\gamma_0=0.02$
and $\Lambda=20\Omega_1$. The choice of the parameters used in for graphical reasons in order to compare
with those plots of the preceding section.}
\label{Fig4}
\end{center}
 \end{figure}

As a further step in our study, we include the interaction with the external reservoir.
In Fig.\ref{Fig4} we can see the behavior of the GP as a function of
the degree of entanglement $\lambda_0$ and the initial angle $\theta_0$ for  $\gamma_0 \neq0$
and $\chi \neq 0$. 
 The presence of a stronger environment will contribute
to the symmetry break and non-monotonic behavior, as can be seen in our case for
a weak environment in Fig.\ref{Fig4}. Once again, we can note that the ``open'' GP
gets a higher value for maximal initial concurrence  and some particular values of
$\theta_0$.  It is easy to note the non-monotonic behavior of the
geometric phase, which exhibits a local maximum and minimum. This characteristic feature
of a energy exchanging process between system and environment (composite in this case) coincides with that shown in Ref.\cite{Hanggi} for a single qubit under a dissipative interaction.
As can be seen,  the stability of phase can be significantly improved via a proper
 choice of the initial state determined by $\theta_0$  which may be crucial for the effectiveness of quantum computation. 
 Another feature that can be noted in Fig.\ref{Fig4} is that, 
as indicated in \cite{Tong2009}, in some cases there is 
 ``saturation" value which is a characteristic value for a given configuration of parameters that 
 do not change much in time.
For example,  an initial angle $\theta_0=\pi/3 \sim 1$, 
 gives a maximum value of GP for a quasi-cyclic period of time. 
In order to study the importance of the initial state, we shall analyze the behavior of the geometric phase for different initial states in the weak coupling limit. 
 
\begin{figure}[h!t]
\begin{center}
\includegraphics[width=8.cm]{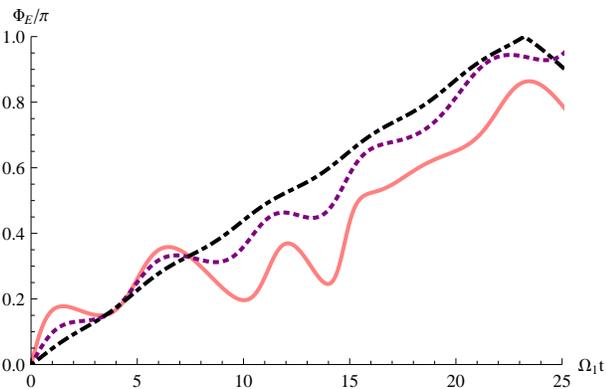}
 \caption{(Color online). The geometric phase for the entangled initial state $\Phi_E$ 
as a function of time. The dashed pink line represents the initial state with $\lambda_0=0.2$;  the dotted dashed purple line 
 $\lambda_0=0.1$ and the black dotted dashed line is for $\lambda_0=0.01$ , all curves for $\theta_0=\pi/5$.
 Parameters used: $\Lambda=20$, $\gamma_0=0.02$ and $\chi=0.1$. The geometric phase is
 normalized with $\pi$.}
\label{Fig5}
\end{center}
 \end{figure}

\begin{figure}[h!t]
\begin{center}
\includegraphics[width=6cm]{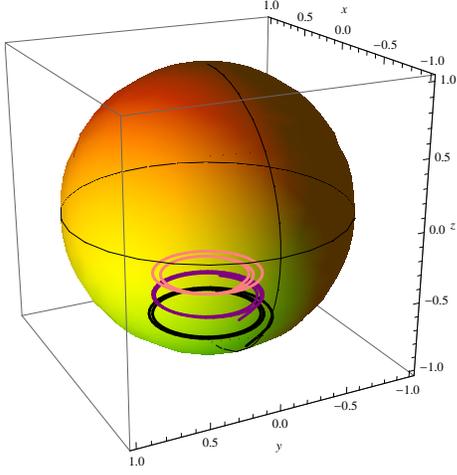}
 \caption{(Color online). The evolution of the system can be illustrated by the path traversed in the
 Bloch sphere. Pink curve for ${\cal {C}}=0.91$, purple curve for ${\cal {C}}=0.71$, black curve  for ${\cal {C}}=0.43$ . Parameters used: $\Lambda=20\Omega_1$, $\gamma_0=0.02$, $\chi=0.1\Omega_1$ and $\theta_0=\pi/5$.}
\label{Fig6}
\end{center}
 \end{figure}

The geometric nature is the main feature of this
phase in unitary closed systems. However, in the open quantum systems framework,
this characteristic is not that evident. There have been many studies for one qubit systems
but none of them has been conclusive about this topic. 
A nice way to see that the change of the GP is associated to the path traversed by the state 
of system is to plot the path traced by the state in the Bloch sphere.
Therefore, Figs. \ref{Fig5} and \ref{Fig6} are presented  to study the  geometric nature of the GP
in open quantum systems. Both figures can be related by noting that states with a higher value of initial concurrence, have a smaller rate of change of the GP in time (Fig. \ref{Fig5}) and a smaller
change in the path traversed (Fig. \ref{Fig6}), at least for small values of $\theta_0$. 
This is consistent
 with the geometric nature of the GP, which might be rephrased as it depends
 on the path traversed by the state of the system and not on the dynamics.  The latter feature can be seen in the Bloch sphere,
as initially bigger values of $\lambda_0$, imply initially smaller radius of the spiral trajectory 
and a smaller change in time of the GP undergoing an external influence. 
For example, if we compare the dotted dashed black curve with the dashed pink one in Fig.\ref{Fig5}, we can see
 that the latter has a smaller rate of change in time than the former one.  Likewise, in Fig.\ref{Fig6} we can note that initially the black
spiral radius is bigger and hence the change of the path and in the geometric phase is notable 
(during the first cycle,  the black one 
 $25\%$ meanwhile the pink trajectory has changed only a $9\%$; the purple one changes a $16\%$). Therefore, we note
 that, not only is it important a proper choice of the initial value of $\theta_0$
 but the value of the initial concurrence plays a crucial role in order to have a robust GP  undergoing an external influence.

\begin{figure}[h!t]
\begin{center}
\includegraphics[width=8.cm]{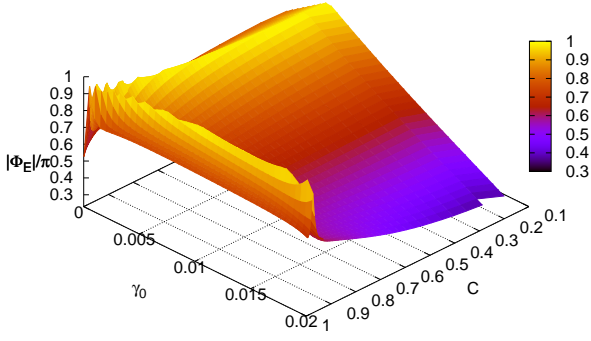}
 \caption{(Color online). The geometric phase, measured in units of $\pi$, as a function of the coupling 
constant to the bosonic bath $\gamma_0$, and the concurrence $\cal C$. Parameters used: $\Lambda=20 \Omega_1$, $\chi=0$, $\theta_0=\pi/3$ and $\tau=2 \pi/\Omega_1$.}
\label{Fig7}
\end{center}
 \end{figure}

In this model we have several environmental parameters such as $\gamma_0$,
$\Lambda$ and $\chi$. We can investigate the effect of the strength of the external bath,
assuming a weak coupling limit, as we show the behavior of the phase as function of the initial 
concurrence and the coupling constant to the external bath.
In Fig.\ref{Fig7}, we can see how this bosonic environment
affects the different initial states of the system. It is easy to note that initially stronger correlations
 seem to contribute to the robustness of the GP in the presence of external couplings $\chi$ and
 $\gamma_0$. The value of $\chi$ adds oscillations to the dynamics of the system.
 Once again, we can note that the stability of the phase can be significantly improved via a proper
 choice of the initial state determined in this case by the value of $\lambda_0$ and $\theta_0$.
 As for this figure, we can enhance
our statement that initially stronger correlated states are prone to being less influenced by
the presence of the environment and hence, the phase change is smaller for those states.
In addition, this figure serves  for a future comparison among the equivalent behavior of the phase if the initial state is a product state as we will show in the next Section. 

It is noteworthy to mention the possibility of implementing 
this model in a NMR-quantum simulator, 
similar to the one used in Ref. \cite{cucchiettiprl}. In this case, we have measured the 
GP-corrections over a qubit induced by another qubit emulating a critical bath. We have obtained this 
correction by measuring the nonunitary evolution of the reduced density matrix of a spin-1/2 
coupled to an environment. The experiments were done using a NMR quantum simulator, where we have 
emulated qualitatively the influence of a critical environment using just a simple one-qubit model.
By adding stochastic fields and further spins, we believe we can quantum-simulate more realistic 
environments and couplings to the system.  With the same idea, it is possible to experimentally test 
the present model in which the entanglement parameter $\lambda_0$ plays an important role. 
In fact, we believe that this configuration will raise a new possibility 
of measuring the GP-corrections implementing a tomography of the reduced density matrix. However,
this experimental test is out of the scope of the present paper.

\subsection{Numerical results for an initial product state}
\label{product}

In this section, we shall see what happens with the geometric phase (set as 
$\Phi_p$) 
of the spin-1/2 particle if the initial state is not entangled,
namely $|\Psi(0)\rangle = |\Psi_1(0) \rangle \times |\Psi_2(0)
\rangle$, where
\begin{eqnarray}
|\Psi_1(0) \rangle &=& \sqrt{1-p} |0_1 \rangle + \sqrt{p} |1_1\rangle \nonumber \\
|\Psi_2(0) \rangle &=& \sqrt{1-q} |0_2 \rangle + \sqrt{q} |1_2\rangle .
\end{eqnarray}
In this case, it is easy to associate the new parameters with the coefficients
$\alpha,\beta,\zeta$ and $\delta$ in order to re-obtain the corresponding new
dynamical equations:
\begin{eqnarray}
 \alpha &=& \sqrt{1-p} \sqrt{1-q} \nonumber  \\
 \beta &=& \sqrt{1-p} \sqrt{q} \nonumber \\
 \zeta &=& \sqrt{p} \sqrt{1-q} \nonumber \\
 \delta &=& \sqrt{p} \sqrt{q} .
\end{eqnarray}

\begin{figure}[h!t]
\begin{center}
\includegraphics[width=9.cm]{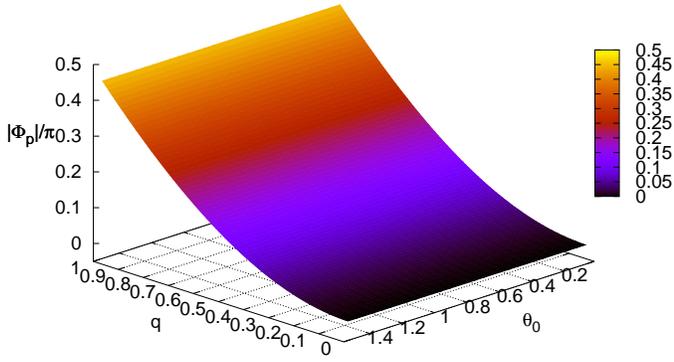}
 \caption{(Color online). Geometric phase for the product initial state $\Phi_p$ as a function of q 
($\lambda_0$) and $\theta_0$, for $\Lambda=20 \Omega_1$, $\chi=0$, and $\gamma_0=0$. }
\label{Fig8}
\end{center}
 \end{figure}

An important consideration is to write the initial state of the bipartite system
with equal parameters as in the entangled case, in order to be able to compare
both cases. Therefore, we can associate $p$ to $\cos^2(\theta_0/2)$
and  write
\begin{eqnarray}
|\Psi_1(0) \rangle &=& \cos(\theta_0/2) |0_1 \rangle + \sin(\theta_0/2) |1_1\rangle \nonumber \\
|\Psi_2(0) \rangle &=& \sqrt{1-q} |0_2 \rangle + \sqrt{q} |1_2\rangle .
\end{eqnarray}
In this way, we can describe the initial state of the system as a function of an angular and radial
coordinates. As before, we can start by considering the case in which the bipartite system is
isolated, that is to say $\gamma_0=\chi=0$. In such a case, we note
that the geometric phase obtained is similar to that of a spin under
unitary evolution $\Phi_g^p= 2 \pi (1-p)$, or equivalently  $\Phi_g^U =2 \pi
\sin^2(\theta_0/2)$, with mod($2\pi$), as shown in Fig.\ref{Fig8}.\\

The next steps in complexity is to consider that both spin-1/2
particles are coupled to each other through the $\chi$ constant and/or to
the external bath through the coupling constant $\gamma_0$. In such cases, the
decoherence factor for the spin 1 (after tracing out spin 2) is
\begin{equation}
 \Gamma_p(t)=e^{-2 \gamma_0 \log{(1+\Lambda^2 t^2)}} \bigg[ \cos(\Omega_R t) + i (2 q -1) \sin(\Omega_R t) \bigg].
\end{equation}
In this case, there are also two eigenvalues, but only one contributes since
$\varepsilon_-(t=0)=0$. So, the
only contribution comes from the eigenvalue $\varepsilon_+$. As the decoherence factor
is complex once again, namely $\Gamma_p(t)= r_(t) e^{i \vartheta_p(t)}$, the computation of the geometric phase is different from the unitary one,
namely
\begin{eqnarray}
 \Phi_p={\rm {arg}} \bigg\{ \sqrt{\varepsilon_+^p(\tau)} \langle\upsilon_+^p(0)| \upsilon_+^p(\tau)\rangle 
  e^{i \varphi_p(t)} \bigg\},
\end{eqnarray}
and 
\begin{eqnarray}
|\upsilon_+^p \rangle &=& \cos \theta_+^p(t) e^{-i \Omega_1 t} e^{i \vartheta_p(t)}|0
\rangle +  \sin \theta_+^p(t),
\end{eqnarray}
with
\begin{eqnarray}
\varphi_p (t) &=& \int_0^{\tau} \bigg( \Omega_1 
-\frac{\partial \vartheta}{dt} \bigg) \cos^2 \theta_+^p(t)~dt \nonumber \\
 \cos \theta_+^p(t) &=& \frac{\sqrt{p(1-p)} r_p(t)}{\sqrt{p(1-p)r_p(t)^2 + p^2 }} \nonumber \\
 \sin \theta_+^p(t) &=& \frac{\sqrt{p} }{\sqrt{p(1-p)r_p(t)^2 + p^2 }}.
\end{eqnarray}

\begin{figure}[h!t]
\begin{center}
\includegraphics[width=7.cm]{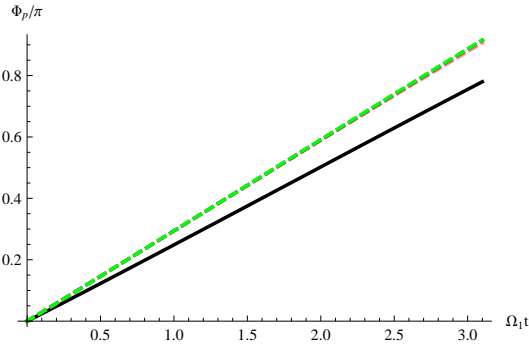}
 \caption{(Color online). Geometric phase as a function of time for different initial states. Dashed lines
 (superposed) represent different values of $q$ and $\theta_0=\pi/5$ and the solid line is for $q=0.05$
 and $\theta_0=\pi/3$. Parameters used:
$\chi=0.1 \Omega_1$, $\Lambda=20 \Omega_1$, $\gamma_0=0.02$.
The choice of the parameters used in for graphical reasons in order to compare
with those plots of the preceding section.}
\label{Fig9}
\end{center}
 \end{figure}

\begin{figure}[h!t]
\begin{center}
\includegraphics[width=7cm]{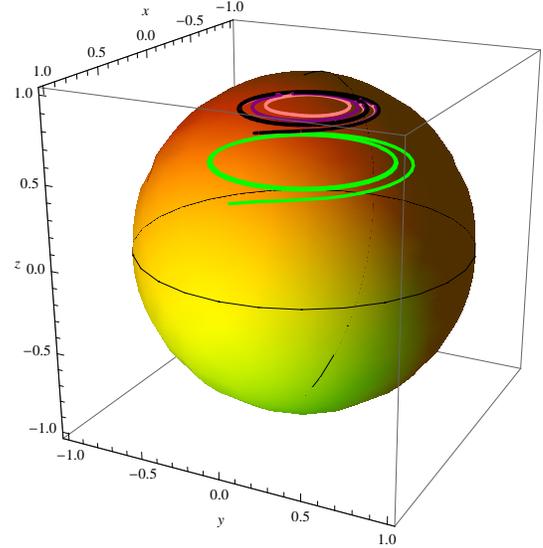}
 \caption{(Color online). Trajectories in the Bloch sphere for the same initial states of Fig.{\ref{Fig6}}.
 Parameters used:
$\chi=0.1\Omega_1$, $\Lambda=20\Omega_1$, $\gamma_0=0.02$.}
\label{Fig10}
\end{center}
 \end{figure}

In order to analyze the importance of the initial state in the robustness of the GP
in open quantum systems, we shall make the same analysis than in the preceding
section. In Fig.\ref{Fig9} we present the GP as a function of the time for different initial
states in the limit of weak coupling with the environment. It is easy to see that the real crucial parameter is the initial value
of $\theta_0$ which sets up the initial state of $|\Psi_1(0)\rangle$.  The same behavior can be seen
in  Fig.\ref{Fig10} for the path traversed in the Bloch sphere. Let's recall that in the preceding
section there was a hierarchy imposed by the value of the initial concurrence.\\

\begin{figure}[h!t]
\begin{center}
\includegraphics[width=7.cm]{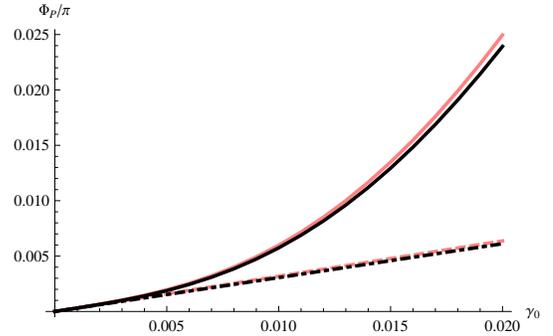}
 \caption{(Color online). The geometric phase, measured in units of $\pi$, for different initial
 states as a function of the dimensionless coupling constant to the bosonic bath $\gamma_0$.  Pink lines
 for $q=0.4$, $\theta=\pi/3$; and black lines  for $q=0.01$, $\theta=\pi/3$. Dotted and dashed lines imply that $\chi=0.1 \Omega_1$ while solid lines imply $\chi=0$. Parameters used: $\Lambda=20 \Omega_1$ and $\tau=2 \pi/\Omega_1$.}
 \label{Fig11}
\end{center}
 \end{figure}

Finally, in Fig.\ref{Fig11} we present the GP for different initial states (the same ones
considered before) as a function of the coupling constant $\gamma_0$, for $\chi=0$
and $\chi \neq 0$. We can see that the presence of the internal dynamics introduced
by $\chi$ reduces considerably the value of the GP in the cases considered.\\

\subsection{The second spin is not coupled to the environment}
\label{nodisipation}

There is another situation that we can study which may result interesting.
It is the case where the spin 2 is not coupled to the external environment,
that means that the spin 1 is coupled to an external environment through the $\lambda_{1n}$
constant and to spin 2 through the $\chi$ constant.
This leads to a redefinition of the decoherence factors that appear in the reduced
density matrix Eq.(\ref{rhogen}). For example, $F_{12}=F_{34}=1$ and $F_{13}=F_{14}=F_{23}=F_{24}=\Gamma_1$
where $\Gamma_1=e^{-2 \gamma_{0} \log(1+\Lambda^2 t^2)}$ is the decoherence factor obtain after tracing out the degrees
of the environment, namely
\begin{equation}
 \Gamma^C(t)= e^{-2 \gamma_{0} \log(1+\Lambda^2 t^2)}((2 \lambda_0 -1) \cos( 2 \chi t)
 - i \sin( 2 \chi t)).
 \nonumber
 \end{equation}
 After tracing out the second spin, we get the reduced density matrix for the
dynamics of spin 1, for the case where the initial state is initially entangled.
\begin{widetext}
\begin{equation}
\tilde{\rho}_{\rm r}(t)=\left(\begin{array}{cc} (\lambda_0 -1/2) \cos(\theta_0) + 1/2
 & 1/2 \sin(\theta_0) e^{-i \Omega_1 t} \Gamma^C(t)  \\
 1/2 \sin(\theta_0) e^{i \Omega_1 t}  \Gamma^{C*}(t)
&  (1/2 -\lambda_0) \cos(\theta_0) + 1/2\\
\end{array}\right).
\label{rho1gensec3}
\end{equation}
\end{widetext}
This expression is very similar to that obtained for the case when the spin 2 is coupled
to the environment in Section \ref{entangled}. However, if we look at the $\Gamma^C(t)$ factor of Eq.(\ref{decoB1})
we see that in the former case there was dissipation introduced by the coupling of the
spin 2 and the environment, while in this case there is not. This is a somewhat reasonable
result since if the spin 2 is effectively coupled to the environment, when we ignore those
degrees of freedom (by tracing them out), we expect to have a loss of energy.
On the other side, just by tracing out the degrees of freedom of one spin, we do not
expect to have a significant loss of energy.
It is important to note that we have chosen to trace out the spin 2 but it is irrelevant
whether we trace spin 1 or spin 2. The results are similar if we study the reduced density
matrix for spin 2 by tracing out the degrees of freedom of spin 1. \\

\section{Conclusions}
\label{IV}

In this article, we have continued with the analysis of the geometric phase in the framework
of a quantum open system, as a natural sequel to some previous works
done in the field. In particular, we have analyzed the geometric phase of a spin
one-half coupled to a noisy composite environment. Initially we have started with a
bipartite state coupled to an external bath and traced out the degrees of freedom
of one of the spins in order to focus on the GP of only one spin. The richness
of the model lays in the fact that the initial bipartite state could be entangled
or not, by introducing the initial entanglement as another free parameter of the
model. We consider it as  a major goal  to understand
the role that the degree of entanglement plays in the dynamics of
the coupled system, as well as in the correction to the geometric phase.

We have calculated the decoherence factor for different initial states of
our system of interest and, consequently, we have analyzed the geometric
phase corrections from the nonunitary evolution.
We have mainly considered two major groups: initial entangled bipartite states
and initial product bipartite states, in order to study the role of the entanglement
in the changes suffer by the geometric phase. In each case, we have computed
 the different decoherence
factors  and numerically obtained the open geometric phase for different physical situations.
 With the help of the study of the geometric nature of the GP, we 
 have shown that the entanglement enhances the sturdiness of the
geometric phase under the presence of an external environment for small values of $\theta_0$.
 States with smaller values of initial entanglement seem to be less robust than others.
 This is related to the fact that the radius of the path traversed is bigger and its curvature
is hence more affected by changes originated by the presence of the environment. 
Another important feature shown is that 
 MES states still remain ``special" or privileged as we have proved in Ref.\cite{PRA81} due to the 
topological nature of the phase of these states.
As shown, in order to implement a measurement of the noise-induced corrections to the GP, it is crucial the choice of initial state. However, we have shown so far that not only
is it important a proper choice of the initial value of $\theta_0$
 but the value of the initial concurrence plays a crucial role in order to have a robust GP  undergoing an external influence. In addition, it is important to state that
 by considering the case in which the temperature $T$ is finite, the main properties
of the geometric phase remain similar. However, it can be expected that the value
of the maximum diminish with increasing temperature.\\

Furthermore, by comparison between the entangled and the product initial bipartite
state, we have seen that 
having an initial entangled bipartite state yields a bigger geometric
phase than in the case of having an initial product bipartite state. In the 
particular case in which the central spin (system of interest) is
completely isolated, it is easy to see that the GP of the initially entangled state
does not correspond
to the geometric phase of an unitarily evolving single spin, i.e.
$\Phi_g^U=\pi(1-\cos\theta_0)$  (due to the initial entanglement). 
However, for an initial product state, 
we have found that the GP, for the uncoupled bipartite system, 
is similar to that of a spin-1/2 under an unitary evolution. This 
 suggests that the degree of entanglement works as an effective 
coupling between the two particles in the 
system, even though one of them is traced out.\\

Finally,  we have noted that the stability of geometric phase with respect 
to decoherence and dissipation is crucial for effectiveness of 
holonomic quantum computation. It is evident that the stability of 
phase can be significantly improved via a proper choice of the initial state determined by the parameters
of the initial state (whether $\lambda_0$ for the entangled state and $\theta_0$ for
the product state). Thus, we consider the better choice an initial entangled bipartite
 state as opposed to a product initial state and with the feasible higher value of concurrence.\\

All in all, we are reporting about a possible scenario where the phase measuring
in an open system is feasible. We claim that the best choice in order to measure 
the open system GP must be take into account the degree of entanglement of the initial 
state. Our model admits the possibility of preparing a MES, which is the best option 
to intent a measure. This conclusion is a byproduct of having this type of model, in 
which one can prepare a two particle initial state, but using just one qubit to 
perform the experiment.

\section{Acknowledgments}

F.C.L was supported by UBA, CONICET,
and ANPCyT, Argentina. PIV acknowledges financial support
from the UNESCO- L'OREAL Women in Science Programme.

\end{document}